\documentstyle[twocolumn,aps,prb,epsfig]{revtex}

\begin{document}
\draft
\twocolumn[\hsize\textwidth\columnwidth\hsize\csname@twocolumnfalse\endcsname%
\title{Time-resolved spectroscopy of multi-excitonic decay in an InAs quantum dot
}
\author{Charles~Santori, Glenn~S.~Solomon\cite{glenadd}, Matthew~Pelton,
        and Yoshihisa~Yamamoto\cite{yamadd}}
\address{Quantum~Entanglement~Project, ICORP, JST,
         E.L.~Ginzton Laboratory, Stanford University,
         Stanford, California 94305}
\date{\today}
\maketitle
\begin{abstract}
The multi-excitonic decay process in a single InAs quantum dot is studied
through high-resolution time-resolved spectroscopy.  A cascaded emission
sequence involving three spectral lines is seen that is described well over
a wide range of pump powers by a simple model.  The measured biexcitonic decay
rate is about 1.5 times the single-exciton decay rate.  This ratio suggests the
presence of selection rules, as well as a significant effect of the Coulomb
interaction on the biexcitonic wavefunction.
\end{abstract}
\vskip2pc]

\narrowtext
Electrostatic interactions play an important role in the energy structures
of semiconductor quantum dots\cite{dots} containing multiple particles.
Although these interactions are usually treated as small perturbations to the
single-particle wavefunctions, they lead to significant energy shifts that
have been measured.\cite{biex1,biex2,biex3,biex4}  These effects
offer possibilities for new quantum-optical devices such as single-photon
sources,\cite{trig1,trig2} entangled photon sources,\cite{benson} and
perhaps even a method to implement quantum logic.\cite{qcomp}

Single-exciton and multi-exciton states, generated by adding electron-hole
pairs to a dot, are of special interest for optical experiments.  They
are ideally the only states that can be generated through resonant optical
excitation of quantum-dot transitions, and they also appear to be the states
most commonly seen in above-band excitation experiments.  Identification
of individual multi-excitonic emission lines was originally based on the
dependence of the emission intensity on the laser excitation power.  More
recently, time-resolved measurements on single quantum dots have become
possible,\cite{roussignol,zwiller,cascade3} and measurements of biexcitonic
emission from single CdSe dots\cite{cascade2} and multi-excitonic
emission from ensembles of CdSe and CuCl nanocrystals\cite{klimov,ikezawa}
and single InAs dots\cite{cascade} have been performed.

Here, we report high-resolution time-resolved measurements on a single
InAs dot using a streak camera system.  After identifying the
single-exciton and multi-exciton emission lines, we measure their decay
rates, and find that the ratio of the one-exciton and biexciton decay
rates is about 1:1.5.  This is closer to the ratio expected for independent
exciton recombination (1:2) than ratios reported for other material
systems.~\cite{cascade2,klimov,ikezawa}  This result suggests the presence of
strong selection rules, as well as a significant effect of the Coulomb interaction
on the wavefunctions of multi-exciton states.  We finally show that the
data may be fit well over a wide range of excitation powers using a simple
model.

\begin{figure}
\centering
\epsfxsize 2.5in
\epsfbox{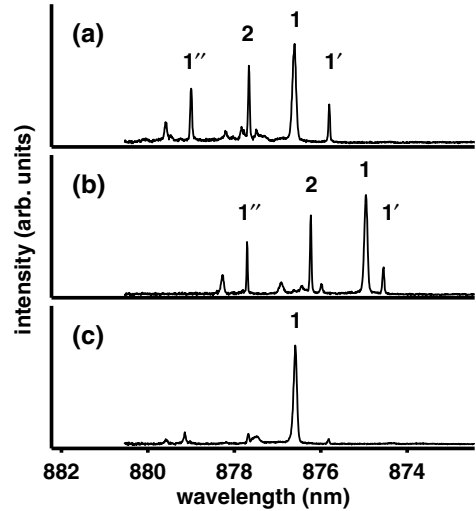}
\vskip.1in
\caption{
Emission spectra of (a) mesa A and (b) mesa B under CW, above-band
($655 \, {\rm nm}$) excitation, and (c) of mesa A under CW, resonant
($858 \, {\rm nm}$) excitation.  Lines 1 and 2 are one- and two-exciton
emission, respectively, while lines $1'$ and $1''$ are believed to
be charged-exciton emission.}
\label{figspectra}
\end{figure}

The InAs self-assembled quantum dots used in this study were grown by
molecular beam epitaxy at a high temperature ($520^{\circ}{\rm C}$) to
increase alloying between the InAs and the surrounding GaAs, yielding
ground-state emission wavelengths in the range of 860-$900\,{\rm nm}$.
The potential wells of the dots are thus rather shallow, and even the
first excited states are close in energy to the wetting layer.  The dots
are approximately $30\,{\rm nm}$ wide, with a density of about
$11\,\mu{\rm m}^{-2}$.  Small mesas (200 or $400\, {\rm nm}$ diameter)
were then fabricated by electron-beam lithography and plasma etching
to isolate single dots.  Mesas containing exactly one dot were identified
through their optical emission spectra.  The spectra shown in
Fig.~\ref{figspectra}(a),(b) were obtained from two mesas (mesas A and B,
respectively) that have similar emission patterns, under continuous-wave (CW)
excitation above the GaAs bandgap ($655\,{\rm nm}$ excitation wavelength).
We identify the lines labeled 1 and 2 as one-exciton and two-exciton emission,
since their dependences on excitation power are linear and quadratic,
respectively, and since photon correlation measurements have confirmed
their link.  The lines labeled $1'$ and $1''$ have linear pump power
dependence, but are identified as charged-state\cite{finley}
emission, since they disappear under excitation resonant with a higher
energy level in the dot, as is seen in Fig.\ \ref{figspectra}(c).
Single mesas at a temperature of $5\,{\rm K}$
were excited from a steep angle ($54^\circ$ from normal) by $3 \, {\rm ps}$
pulses every $13 \, {\rm ns}$ from a Ti-sapphire laser, focused down
to an $18 \, \mu{\rm m}$ effective spot diameter.  The resulting emission
was collected by an ${\rm NA} = 0.5$ aspheric lens, spectrally filtered
to reject scattered laser light, and imaged onto a removable pinhole,
which selected emission from a $5 \, \mu{\rm m}$ region of the sample.
The emission was then sent to an alignment camera, a spectrometer, or a
streak camera system, which included a monochrometer that determined
both the spectral ($0.13 \, {\rm nm}$) and temporal ($25 \, {\rm ps}$)
resolutions.  The streak camera recorded the emission following the
excitation pulses, averaged over about 5 minutes ($2.3 \times 10^{10}$
pulses).  The resulting images were corrected for dark current, non-uniform
sensitivity, and a small number of cosmic ray events.
\begin{figure}
\centering
\epsfxsize 3.2in
\epsfbox{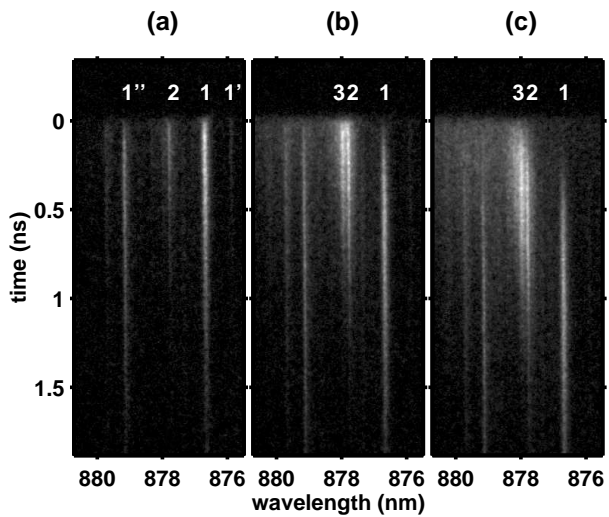}
\vskip.1in
\caption{
Streak camera images of emission from mesa A under pulsed, $708 \, {\rm nm}$
laser excitation with powers (a) 27 $\mu{\rm W}$, (b) 108 $\mu{\rm W}$,
and (c) 432 $\mu{\rm W}$.  For larger powers, a cascaded emission
is seen, with the multi-excitonic emission (lines 2, 3) occurring first,
followed by one-exciton emission (line 1).}
\label{figimages}
\end{figure}
Images obtained under above-band ($708 \, {\rm nm}$), pulsed laser excitation of
mesa A with three different excitation powers are shown in Fig.\ \ref{figimages}.
The observed emission lines are labeled as in Fig.\ \ref{figspectra}.
Figure\ \ref{figimages}(a) shows that under weak excitation ($27 \, \mu{\rm W}$),
the single-exciton line (line 1) appears less than $0.1 \, {\rm ns}$ after the
excitation pulse, and then decays exponentially.  We attribute the small initial
delay to the time required for electrons and holes generated by the excitation pulse
to be captured by the dot.  However, when the excitation power is increased to
$108 \, \mu{\rm W}$, Fig.\ \ref{figimages}(b) shows that line 1 reaches its maximum
only after a delay of about $0.5 \, {\rm ns}$.  Most of the emission
immediately after the excitation pulse now comes from the multi-exciton lines
2 and~3.  In this case, the laser pulse initially creates several electron-hole
pairs, and some time is required before the population in the dot reduces to one
electron-hole pair, after which one-exciton emission occurs.  Under strong
excitation ($432 \, \mu{\rm W}$), one can see from Fig.\ \ref{figimages}(c) that
the delay in the one-exciton emission is even longer, and a delay also appears
for lines 2~and~3.  Only the broadband emission in the vicinity of the
multi-exciton lines is seen to appear immediately after the excitation pulse.

The multi-excitonic decay process may be described by the following rate
equation:
\begin{equation}
\frac{d}{dt}P_n(t) = \gamma_{n+1} P_{n+1}(t) - \gamma_n P_n(t) \, , \\ \label{rateeq}
\end{equation}
where $P_n(t)$ is the probability that $n$ electron-hole pairs exist in the
dot at time $t$, and $\gamma_n$ is the decay rate of the $n$-pair state.
The creation of charged dot states is not considered here.  Instead,
we apply this model only to neutral-dot outcomes following an excitation
pulse by excluding emission lines $1'$ and $1''$ from the analysis, and noting
that radiative decay beginning with a neutral state cannot generate charged
states.  The form of $\gamma_n$ depends strongly on the nature of the states of the
system.  For an uncorrelated system with no recombination selection rules, one might
expect $\gamma_n = n^2 \gamma_1$.  For a dot much smaller than
the exciton Bohr radius, the Coulomb interaction has little effect on
the wavefunctions of multi-exciton states, and due to selection rules
one expects approximately independent recombination, $\gamma_n = n \gamma_1$.
For a larger dot, it has been predicted that the Coulomb interaction
produces a spatial separation between holes in multi-exciton states,
so that, for example, the biexciton state resembles a
molecule \cite{cascade2,takagahara}.  In this case, one
expects $\gamma_n < n \gamma_1$, and this has been observed for
CdSe quantum dots.\cite{cascade2}

\begin{figure}
\centering
\epsfxsize 2.7in
\epsfbox{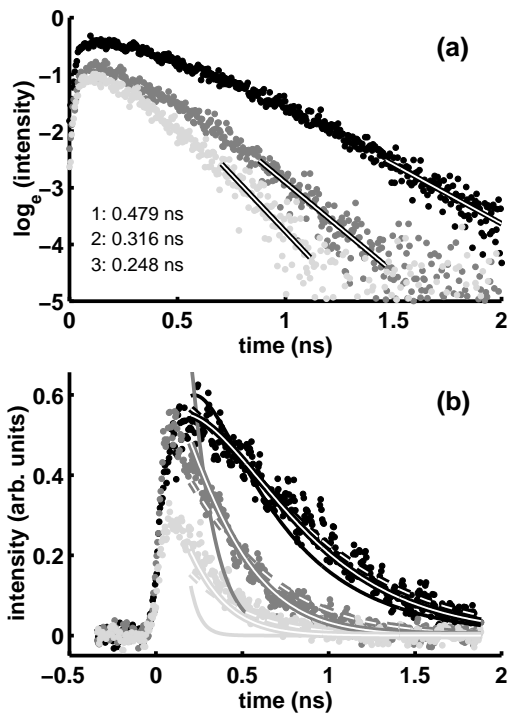}
\vskip.1in
\caption{
(a) Time-dependent intensities of lines 1 (black), 2 (dark gray), and
3 (light gray) from mesa A under pulsed, resonant ($858 \, {\rm nm}$,
$2 \,{\rm mW}$) laser excitation.  Exponential fits are applied after sufficient
decay has occurred to estimate the decay times.  (b)  Time-dependent intensities
under above-band ($708 \, {\rm nm}$, $54 \, \mu{\rm W}$) excitation, compared to
three models: $\gamma_n = n \gamma_1$ (hollow lines), measured
decay rates (hollow, dashed), and $\gamma_n = n^2 \gamma_1$ (solid).}
\label{fignew}
\end{figure}

The decay rates $\gamma_n$ in Eq.~\ref{rateeq} can be measured directly from
the time-dependent intensities of the lines corresponding to $P_n(t)$, when
$P_n(t) \gg P_{n+1}(t)$.  To perform this measurement as accurately as
possible, we tuned the excitation laser to a resonance at about
$858 \, {\rm nm}$, as in Fig.\ \ref{figspectra}(c), creating electron-hole
pairs directly inside the dot, which rapidly
relax to a lowest-energy state.  This way, the delayed capture of electrons
and holes, which can alter the apparent decay rates, does not occur, as it
could in the above-band excitation case.  The excitation power (about $2 \, {\rm mW}$)
was chosen so that multi-exciton states were created with significant
probability.  The intensities of lines 1, 2, and 3 were calculated in a
straightforward manner, by integrating the streak camera image over strips
about $0.2 \, {\rm nm}$ wide, defined to include all of the emission from each line.
Lines 2 and 3 were not completely resolved, and the integration boundary was
placed midway between them.  A further concern for the multi-exciton
lines 2 and 3 is that any weak background emission related to the one-exciton
or charged-exciton states having a slower decay will cause
significant distortion at large $t$, which is where we must measure
the decay rates.  With these cautions in mind, Fig.\ \ref{fignew}(a) shows
the time-dependent intensities of lines 1, 2, and 3 under resonant excitation
and a 60-minute integration, plotted on a semi-logarithmic scale.
The slopes are estimated over the indicated regions by
least-squares exponential fits.  The time constants
obtained are $0.479 \, {\rm ns}$, $0.316 \, {\rm ns}$, and $0.248 \, {\rm ns}$,
respectively.  The one-exciton (line 1) lifetime seen here is close to the
value of $0.47 \, {\rm ns}$ we have measured with weaker excitation powers.  We obtain
$\gamma_2 = 1.52 \gamma_1$, a result in between the small-dot limit
($\gamma_2 = 2 \gamma_1$) and what has been reported for CdSe dots
($\gamma_2 \approx \gamma_1$).\cite{cascade2}  This is consistent
with the presence of selection rules and a departure from the
small-dot limit, due to the influence of the Coulomb interaction on
the multi-exciton wavefunctions.   Line 3 is likely due to
emission from the 3-exciton state, since its wavelength relative to the
one-exciton and biexciton lines is similar to what has been reported
elsewhere for the 3-exciton line,\cite{hartman} and this identification
is made plausible here by its time dependence.  With this assumption,
we obtain $\gamma_3 = 1.93 \gamma_1$.
We now wish to model the multi-excitonic decay process under above-band excitation
over a wide range of excitation powers, to show that the behavior of lines
1, 2, and 3 is consistent with a cascaded decay.  We assume that photons from
the laser excitation pulse are absorbed independently by the GaAs surrounding
the dot to form electron-hole pairs, and that these pairs are then independently
captured by the dot, so that the initial population of the dot follows a
Poisson distribution with mean $\mu$.  The dot then decays according to
Eq.~\ref{rateeq}, and we assume that the observed intensity from the
$n$-exciton state is $I_n = \gamma_n I_0 P_n$, where $I_0$ includes the
collection efficiency.  We make this assumption for simplicity, though it
does not take into account the presence of multiple emission lines for $n>2$.
Fig.\ \ref{fignew}(b) shows the time-dependent intensities of lines 1, 2, and
3 under above-band ($708 \, {\rm nm}$, $54 \, {\rm mW}$) excitation, along
with three models, each simultaneously fit to lines 1 and 2.
The predicted 3-exciton behavior is also shown, in comparison to
line 3.

In the first model (hollow lines), we assume independent recombination,
$\gamma_n = n \gamma_1$.  Although this assumption differs substantially from the
measured rates, its simplicity is appealing.  The resulting time-dependent
probabilities have the simple form of a Poisson distribution with exponentially
decaying mean:
\begin{equation}
P_n(t) = (\mu e^{-\gamma_1 t})^n \exp(- \mu e^{-\gamma_1 t}) / n! \, , \label{poiseq}
\end{equation}
where $\mu$ is again the mean initial exciton number.  This model is also well
suited to handle an additional complication noticeable in the data.
For above-band excitation, a finite time is required for the dot to capture
the excitons generated by the laser pulse, with some recombination occurring
during this time.  But since, in this model, the excitons are both generated
and annihilated independently, a Poisson distribution always holds, and it is
sufficient to wait until the capture process has
finished ($0.2 \, {\rm ns}$) to begin fitting the data to Eq.~\ref{poiseq}.
In the second model (hollow, dashed lines), Eq.~\ref{rateeq} is solved
numerically, using the measured decay rates and assuming a Poisson distribution,
beginning at $0.2 \, {\rm ns}$.  The rates $\gamma_n$ for $n > 3$ had to be extrapolated 
from the trend seen for $n \le 3$, but have little effect on the result for
this excitation power.  In the third model (thin solid lines), Eq.~\ref{rateeq}
is solved numerically assuming $\gamma_n = n^2 \gamma_1$, as one would
expect with no recombination selection rules.  For all three models, only two fitting
parameters are used, the initial mean number of excitons $\mu$, and the
collection efficiency constant, $I_0$.  From these two fitting parameters, all
three curves (one-exciton, biexciton, 3-exciton) are obtained
simultaneously.  Two of these cascaded decay models, the model with independent
exciton decay ($\gamma_n = n \gamma_1$) and the model using measured decay
rates, fit the data reasonably well (mean squared errors $8.2 \times 10^{-4}$
and $9.4 \times 10^{-4}$, respectively).  The other model ($\gamma_n = n^2 \gamma_1$)
fits the data poorly (mean squared error $4.4 \times 10^{-3}$).

\begin{figure}
\centering
\epsfxsize 3.2in
\epsfbox{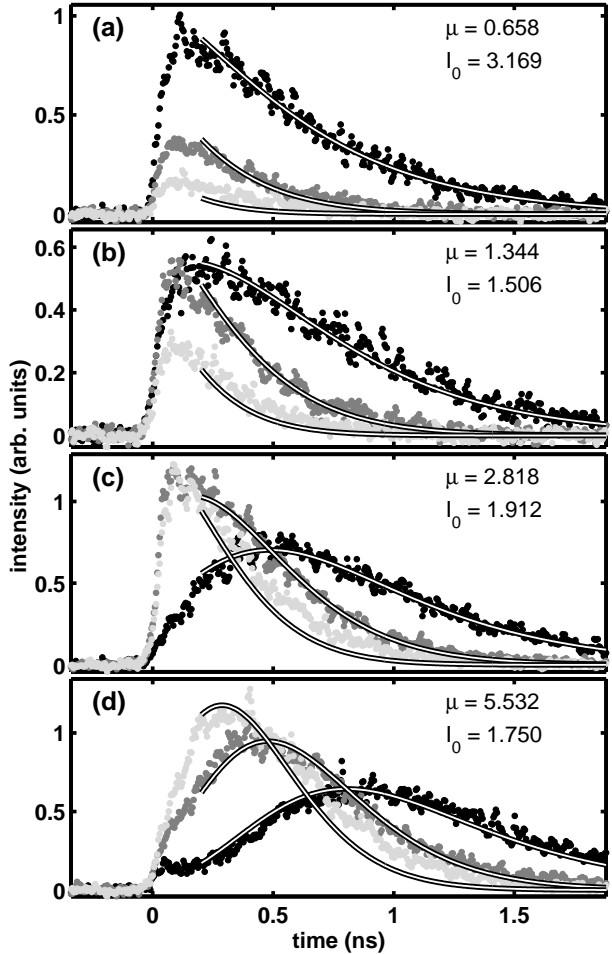}
\vskip.1in
\caption{
Time-dependent intensities of lines 1 (black), 2 (dark gray), and
3 (light gray) from mesa A under pulsed, $708 {\rm nm}$ laser
excitation with powers (a) 27 $\mu{\rm W}$, (b) 54 $\mu{\rm W}$,
(c) 108 $\mu{\rm W}$, and (d) 432 $\mu{\rm W}$.  Hollow lines
show model fit results, and fit parameters are given at upper-right.}
\label{figfits}
\end{figure}

To demonstrate that lines 1, 2, and 3 are well described by a cascaded emission
process, we fit the simplest model ($\gamma_n = n \gamma_1$) to the data for
four different pump powers in Fig.\ \ref{figfits}.  The value of
$\gamma_1$ was fixed at $(0.47 \, {\rm ns})^{-1}$, and the obtained values of
the fitting parameters $I_0$ and $\mu$ are shown.  The value of $\mu$ increases
linearly with excitation power at first, and then begins to saturate.
Ideally, $I_0$ should be the same for each streak camera image, but in our case,
we had to fit it separately for each image due largely to spatial sample drift
and streak-camera gain drift.  The fit was performed to minimize the combined
errors for lines 1 and 2.
For all three lines, the simple model provides excellent
agreement with the data in the weak, moderate, and strong-excitation cases,
supporting the presence of a multi-excitonic decay sequence involving these
lines.
In summary, we have observed multi-excitonic decay spectra from a single quantum
dot using a streak camera, providing high temporal resolution.  We have measured
the decay rates of several lines, and found that the biexciton decay rate is about
1.5 times the one-exciton decay rate, suggesting strong recombination selection
rules and a significant influence of the Coulomb interaction on the multi-exciton
wavefunctions.  We then showed that, under above-band excitation, the time dependence
of the emission lines is well described over a wide range of excitation powers by
a simple model for cascaded emission.

Financial assistance for C.~S. was provided by the National Science
Foundation.  Financial assistance for C.~S. and M.~P. was provided by Stanford
University.  G.~S.~S. is partially supported by ARO (D. Woolard).


\begin{references}
%
\bibitem[*]{glenadd}
Also at Solid-State Photonics Laboratory, Stanford University, Stanford, CA.
%
\bibitem[\dagger]{yamadd}
Also at NTT Basic Research Laboratories, Atsugishi, Kanagawa, Japan.
%
\bibitem{dots}
D.~Bimberg, M.~Grundmann, and N.~N.~Ledentsov, {\it Quantum Dot Heterostructures}
(John Wiley \& Sons, Chichester, 1999).
%
\bibitem{biex1}
A.~Kuther, M.~Bayer, A.~Forchel, A.~Gorbunov, V.~B.~Timofeev, F.~Sch\"{a}fer, and
J.~P.~Reithmaier, Phys. Rev. B {\bf 58}, R7508 (1998).
%
\bibitem{biex2}
H.~Kamada, H.~Ando, J.~Temmyo, and T.~Tamamura, Phys. Rev. B {\bf 58}, 16243 (1998).
%
\bibitem{biex3}
L.~Landin, M.~S.~Miller, M.-E.~Pistol, C.~E.~Pryor, and L.~Samuelson,
Science {\bf 280}, 262 (1998).
%
\bibitem{biex4}
M.~Bayer, O.~Stern, P.~Hawrylak, S.~Fafard, and A.~Forchel, Nature (London) {\bf 405}, 923 (2000).
%
\bibitem{trig1}
C.~Santori, M.~Pelton, G.~Solomon, Y.~Dale, and Y.~Yamamoto, Phys. Rev. Lett.
{\bf 86}, 1502 (2001).
%
\bibitem{trig2}
P.~Michler, A.~Kiraz, C.~Becher, W.~V.~Schoenfeld, P.~M.~Petroff, L.~Zhang, E.~Hu, and
A.~Imamo\u{g}lu, Science {\bf 290}, 2282 (2000).
%
\bibitem{benson}
O.~Benson, C.~Santori, M.~Pelton, and Y.~Yamamoto, Phys. Rev. Lett. {\bf 84},
2513 (2000).
%
\bibitem{qcomp}
F.~Troiani, U.~Hohenester, and E.~Molinari, Phys. Rev. B {\bf 62}, R2263 (2000).
%
\bibitem{roussignol}
Ph.~Roussignol, W.~Heller, A.~Filoramo, and U.~Bockelmann, Physica E {\bf 2}, 588 (1998).
%
\bibitem{zwiller}
V.~Zwiller, M.-E.~Pistol, D.~Hessman, R.~Cederstr\"{o}m, W.~Seifert, and L.~Samuelson,
Phys. Rev. B {\bf 59}, 5021 (1999).
%
\bibitem{cascade3}
D.~V.~Regelman, E.~Dekel, D.~Gershoni, W.~V.~Schoenfeld, and P.~M.~Petroff,
Phys. Stat. Sol. (b), {\bf 224}, 343 (2001).
%
\bibitem{cascade2}
G.~Bacher, R.~Weigand, J.~Seufert, V.~D.~Kulakovskii, N.~A.~Gippius, A.~Forchel, K.~Leonardi,
and D.~Hommel, Phys. Rev. Lett. {\bf 83}, 4417 (1999).
%
\bibitem{klimov}
V.~I.~Klimov, A.~A.~Mikhailovsky, D.~W.~McBranch, C.~A.~Leatherdale,
and M.~G.~Bawendi, Science {\bf 287}, 1011 (2000).
%
\bibitem{ikezawa}
M.~Ikezawa and Y.~Masumoto, Phys. Rev. B {\bf 53}, 13694 (1996).
%
\bibitem{cascade}
E.~Dekel, D.~V.~Regelman, D.~Gershoni, E.~Ehrenfreund, W.~V.~Shoenfeld,
and P.~M.~Petroff, Phys. Rev. B {\bf 62}, 11038 (2000).
%
\bibitem{finley}
J.~J.~Finley, P.~W.~Fry, A.~D.~Ashmore, A.~Lema\^{\i}tre, A.~I.~Tartakovskii, R.~Oulton,
D.~J.~Mowbray, M.~S.~Skolnick, M.~Hopkinson, P.~D.~Buckle, and P.~A.~Maksym,
Phys. Rev. B {\bf 63}, 161305(R) (2001).
%
%
\bibitem{takagahara}
T. Takagahara, Phys. Rev. B {\bf 39}, 10206 (1989).
%
\bibitem{hartman}
A.~Hartmann, Y.~Ducommun, E.~Kapon, U.~Hohenester, and E.~Molinari,
Phys. Rev. Lett. {\bf 84}, 5648 (2000). 
%
\end{references}
\end{document}